\documentstyle[12pt,epsfig]{article}

\newskip\humongous \humongous=0pt plus 1000pt minus 1000pt

\newif\ifdtup

\def\pr#1{#1^\prime}

\def\beq{\begin{equation}}
\def\eeq{\end{equation}}

\def\beqn{\begin{eqnarray}}
\def\eeqn{\end{eqnarray}}
\relax

\def\dotx{\dotx{\dot\overline{x}}}

\relax

\catcode`\@=11

\@addtoreset{equation}{section}
\def\theequation{\thesection\arabic{equation}}

\def\@normalsize{\@setsize\normalsize{15pt}\xiipt\@xiipt
\abovedisplayskip 14pt plus3pt minus3pt%
\belowdisplayskip \abovedisplayskip
\abovedisplayshortskip \z@ plus3pt%
\belowdisplayshortskip 7pt plus3.5pt minus0pt}

\def\small{\@setsize\small{13.6pt}\xipt\@xipt
\abovedisplayskip 13pt plus3pt minus3pt%
\belowdisplayskip \abovedisplayskip
\abovedisplayshortskip \z@ plus3pt%
\belowdisplayshortskip 7pt plus3.5pt minus0pt
\def\@listi{\parsep 4.5pt plus 2pt minus 1pt
     \itemsep \parsep
     \topsep 9pt plus 3pt minus 3pt}}

\@twosidetrue
\relax

\catcode`@=12

\catcode`\@=11

\def\section{\@startsection{section}{1}{\z@}{3.5ex plus 1ex minus
   .2ex}{2.3ex plus .2ex}{\large\bf}}

\def\thesection{\arabic{section}.}

\def\appendix{\setcounter{section}{0}
 \def\thesection{APPENDIX \Alph{section}:}
 \def\theequation{\Alph{section}.\arabic{equation}}}
\def\ps@headings{\def\@oddfoot{}\def\@evenfoot{}
\def\@oddhead{\hbox{}\hfill
 \makebox[.5\textwidth]{\raggedright\ignorespaces --\thepage{}--
 \hfill {}}}  %instead of {\rm FERMILAB--Pub--\FERMIPUB}}}
\def\@evenhead{\@oddhead}
\def\subsectionmark##1{\markboth{##1}{}}
}

\ps@headings

\catcode`\@=12

\def\figcap{\section*{Figure Captions\markboth
 {FIGURECAPTIONS}{FIGURECAPTIONS}}\list
 {Fig. \arabic{enumi}:\hfill}{\settowidth\labelwidth{Fig. 999:}
 \leftmargin\labelwidth
 \advance\leftmargin\labelsep\usecounter{enumi}}}
 \relax
\def\tablecap{\section*{Table Captions\markboth
 {TABLECAPTIONS}{TABLECAPTIONS}}\list
 {Table \arabic{enumi}:\hfill}{\settowidth\labelwidth{Table 999:}
 \leftmargin\labelwidth
 \advance\leftmargin\labelsep\usecounter{enumi}}}
 \relax
\def\reflist{\section*{References\markboth
 {REFLIST}{REFLIST}}\list
 {[\arabic{enumi}]\hfill}{\settowidth\labelwidth{[999]}
 \leftmargin\labelwidth
 \advance\leftmargin\labelsep\usecounter{enumi}}}
 \relax

\catcode`\@=11

\def\ps@headings{\def\@oddfoot{}\def\@evenfoot{}
\def\@oddhead{\hbox{}\hfill
 \makebox[.5\textwidth]{\raggedright\ignorespaces --\thepage{}--
 \hfill {}}}    %instead of {\rm FERMILAB--Pub--\FERMIPUB}}}
\def\@evenhead{\@oddhead}
\def\subsectionmark##1{\markboth{##1}{}}
}

\ps@headings

\catcode`\@=12

\relax

\catcode`\@=11
\def\prm{\fam \z@}
\catcode`\@=12

\relax
\def\pl#1#2#3{{\it Phys. Lett. }{\bf #1}(19#2)#3}
\def\zp#1#2#3{{\it Z. Phys. }{\bf #1}(19#2)#3}
\def\prl#1#2#3{{\it Phys. Rev. Lett. }{\bf #1}(19#2)#3}

\def\pr#1#2#3{{\it Phys. Rev. }{\bf #1}(19#2)#3}
\def\np#1#2#3{{\it Nucl. Phys. }{\bf #1}(19#2)#3}

\def    \hepph  #1 {{\tt hep-ph/#1}}
\def    \hepex  #1 {{\tt hep-ex/#1}}

\relax

  \newcommand{\ccaption}[2]{
    \begin{center}
    \parbox{0.85\textwidth}{
      \caption[#1]{\small{\it{#2}}}
      }
    \end{center}
    }
\begin{document}            
\newcommand\sss{\scriptscriptstyle}
\newcommand\mug{\mu_\gamma}
\newcommand\mue{\mu_e}
\newcommand\muf{\mu_{\sss F}}
\newcommand\mur{\mu_{\sss R}}
\newcommand\muo{\mu_0}
\newcommand\me{m_e}
\newcommand\as{\alpha_{\sss S}}         
\newcommand\ep{\epsilon}
\newcommand\Th{\theta}
\newcommand\epb{\overline{\epsilon}}
\newcommand\aem{\alpha_{\rm em}}
\newcommand\refq[1]{$^{[#1]}$}
\newcommand\avr[1]{\left\langle #1 \right\rangle}
\newcommand\lambdamsb{\Lambda_5^{\rm \sss \overline{MS}}}
\newcommand\qqb{{q\overline{q}}}
\newcommand\qb{\overline{q}}
\newcommand\MSB{{\rm \overline{MS}}}
\newcommand\DIG{{\rm DIS}_\gamma}
\newcommand\CA{C_{\sss A}}
\newcommand\DA{D_{\sss A}}
\newcommand\CF{C_{\sss F}}
\newcommand\TF{T_{\sss F}}
\newcommand\Jetlist{\{J_l\}_{1,3}}
\newcommand\aoat{\sss a_{\sss 1} a_{\sss 2}}
\newcommand\SFfull{\{k_l\}_{1,6}}
\newcommand\SFfullbi{\{k_l\}_{1,6}^{[i]}}
\newcommand\SFTfull{\{k_l\}_{1,5}}
\newcommand\SFj{\{k_l\}_{3,6}}
\newcommand\FLfull{\{a_l\}_{1,6}}
\newcommand\FLfullbi{\{a_l\}_{1,6}^{[i]}}
\newcommand\FLTfull{\{a_l\}_{1,5}}
\newcommand\FLFj{\{a_l\}_{3,6}}
\newcommand\SFjbi{\{k_l\}_{3,6}^{[i]}}
\newcommand\FLjbi{\{a_l\}_{3,6}^{[i]}}
\newcommand\SFjexcl{\{k_l\}_{i,j}^{[np..]}}
\newcommand\SCollfull{\{k_l\}_{1,7}^{[ij]}}
\newcommand\SColl{\{k_l\}_{3,7}^{[ij]}}
\newcommand\FLCollfull{\{a_l\}_{1,7}^{[ij]}}
\newcommand\FLColl{\{a_l\}_{3,7}^{[ij]}}
\newcommand\STj{\{k_l\}_{3,5}}
\newcommand\FLTj{\{a_l\}_{3,5}}
\newcommand\FLNj{\{a_l\}_{3}^{N+2}}
\newcommand\FLNmoj{\{a_l\}_{3}^{N+1}}
\newcommand\FLNfullj{\{a_l\}_{1}^{N+2}}
\newcommand\FLNmofullj{\{a_l\}_{1}^{N+1}}
\newcommand\Argfull{\FLfull;\SFfull}
\newcommand\ArgTfull{\FLTfull;\SFTfull}
\newcommand\KtoKF{(k_1,k_2\to\SFj\,;\FLFj)}
\newcommand\KtoKT{(k_1,k_2\to\STj\,;\FLTj)}
\newcommand\FLsum{\sum_{\{a_l\}}}
\newcommand\FLFsum{\sum_{\FLFj}}
\newcommand\FLFsumbi{\sum_{\FLjbi}}
\newcommand\FLTsum{\sum_{\FLTj}}
\newcommand\FLNsum{\sum_{\FLNj}}
\newcommand\FLNmosum{\sum_{\FLNmoj}}
\newcommand\MF{{\cal M}^{(4)}}
\newcommand\MT{{\cal M}^{(3)}}
\newcommand\MTz{{\cal M}^{(3,0)}}
\newcommand\MTo{{\cal M}^{(3,1)}}
\newcommand\MTi{{\cal M}^{(3,i)}}
\newcommand\MTmn{{\cal M}^{(3,0)}_{mn}}
\newcommand\MN{{\cal M}^{(N)}}
\newcommand\MNmo{{\cal M}^{(N-1)}}
\newcommand\MNmoij{{\cal M}^{(N-1)}_{ij}}
\newcommand\MNmoV{{\cal M}^{(N-1,v)}}
\newcommand\MFsj{\MF(k_1,k_2\to\SFj)}
\newcommand\MTsj{\MT(k_1,k_2\to\STj)}
\newcommand\MTisj{\MTi(k_1,k_2\to\STj)}
\newcommand\PHIFsj{\phi_4(k_1,k_2\to\SFj)}
\newcommand\PHITsj{\phi_3(k_1,k_2\to\STj)}
\newcommand\uoffct{\frac{1}{4!}}
\newcommand\uotfct{\frac{1}{3!}}
\newcommand\uoNfct{\frac{1}{N!}}
\newcommand\uoNmofct{\frac{1}{(N-1)!}}
\newcommand\uoxic{\left(\frac{1}{\xi}\right)_c}
\newcommand\uoxiic{\left(\frac{1}{\xi_i}\right)_c}
\newcommand\uoxilc{\left(\frac{\log\xi}{\xi}\right)_c}
\newcommand\uoxiilc{\left(\frac{\log\xi_i}{\xi_i}\right)_c}
\newcommand\uoyim{\left(\frac{1}{1-y_i}\right)_+}
\newcommand\uoyimdi{\left(\frac{1}{1-y_i}\right)_{\delta_{\sss I}}}
\newcommand\uoyimpdi{\left(\frac{1}{1\mp y_i}\right)_{\delta_{\sss I}}}
\newcommand\uoyjmdo{\left(\frac{1}{1-y_j}\right)_{\delta_o}}
\newcommand\uoyip{\left(\frac{1}{1+y_i}\right)_+}
\newcommand\uoyipdi{\left(\frac{1}{1+y_i}\right)_{\delta_{\sss I}}}
\newcommand\uoyilm{\left(\frac{\log(1-y_i)}{1-y_i}\right)_+}
\newcommand\uoyilp{\left(\frac{\log(1+y_i)}{1+y_i}\right)_+}
\newcommand\uozm{\left(\frac{1}{1-z}\right)_+}
\newcommand\uozlm{\left(\frac{\log(1-z)}{1-z}\right)_+}
\newcommand\SVfact{\frac{(4\pi)^\ep}{\Gamma(1-\ep)}
                   \left(\frac{\mu^2}{Q^2}\right)^\ep}
\newcommand\gs{g_{\sss S}}
\newcommand\Icol{\{d_l\}}
\newcommand\An{{\cal A}^{(n)}}
\newcommand\Mn{{\cal M}^{(n)}}
\newcommand\Nn{{\cal N}^{(n)}}
\newcommand\Anu{{\cal A}^{(n-1)}}
\newcommand\Mnu{{\cal M}^{(n-1)}}
\newcommand\Sumae{\sum_{d_{e}}}
\newcommand\Sumaecl{\sum_{d_{e},\Icol}}
\newcommand\Sumaeae{\sum_{d_{e},d_{e}^{\prime}}}
\newcommand\Sumhe{\sum_{h_{e}}}
\newcommand\Sumhep{\sum_{h_{e}^\prime}}
\newcommand\Sumhehe{\sum_{h_{e},h_{e}^{\prime}}}
\newcommand\Pgghhh{S_{gg}^{h_e h_i h_j}}
\newcommand\Pggplus{S_{gg}^{+ h_i h_j}}
\newcommand\Pggminus{S_{gg}^{- h_i h_j}}
\newcommand\Pgghphh{S_{gg}^{h_e^\prime h_i h_j}}
\newcommand\Pqgplus{S_{qg}^{+ h_i h_j}}
\newcommand\Pqgminus{S_{qg}^{- h_i h_j}}
\newcommand\Pgqplus{S_{gq}^{+ h_i h_j}}
\newcommand\Pgqminus{S_{gq}^{- h_i h_j}}
\newcommand\Pqqplus{S_{qq}^{+ h_i h_j}}
\newcommand\Pqqminus{S_{qq}^{- h_i h_j}}
\newcommand\Hnd{\{h_l\}}
\newcommand\LC{\stackrel{\sss i\parallel j}{\longrightarrow}}
\newcommand\LCu{\stackrel{\sss 1\parallel j}{\longrightarrow}}
\newcommand\Physvar{\{V_l\}}
\newcommand\Physcm{\{\bar{V}_l\}}
\newcommand\Partvar{\{v_l\}}
\newcommand\Partcm{\{\bar{v}_l\}}
\newcommand\Partvarcm{\{v_l(\bar{v})\}}
\newcommand\Partcmset{\{\bar{v}_l^{(i)}\}}
\newcommand\Partvarcmset{\{v_l(\bar{v}^{(i)})\}}
\def\A#1#2{\la#1#2\ra} 
\def\B#1#2{[#1#2]} 
\newcommand{\la}{\langle}
\newcommand{\ra}{\rangle}
\newcommand{\nn}{\nonumber}
\newcommand\Qb{\overline{Q}}
\renewcommand\topfraction{1}       % Max. Fraz. di pagina per float in t
\renewcommand\bottomfraction{1}    % Max. Fraz. di pagina per float in b
\renewcommand\textfraction{0}      % Min. Fraz. di pagina per testo 
\setcounter{topnumber}{5}          % Max # float in position t
\setcounter{bottomnumber}{5}       % Max # float in position b
\setcounter{totalnumber}{5}        % Max # float in same page
\setcounter{dbltopnumber}{2}       % Max # large float
\newsavebox\tmpfig
\newcommand\settmpfig[1]{\sbox{\tmpfig}{\mbox{\ref{#1}}}}
%
%%%%%%%%%input files
%\input mcjettitle
\begin{titlepage}
\nopagebreak
{\flushright{
        \begin{minipage}{4cm}
        ETH-TH/97-14 \hfill \\
        hep-ph/9706545\hfill \\
        \end{minipage}        }

}
\vfill
\begin{center}
{\LARGE { \bf \sc A general approach to\\[0.3cm]
                  jet cross sections in QCD}} 
\vskip .5cm
{\bf Stefano FRIXIONE}\footnote{Work supported by the Swiss National
Foundation.}
\\                    
\vskip .1cm
{Theoretical Physics, ETH, Zurich, Switzerland} \\
\end{center}
\nopagebreak
\vfill
%\vskip 3cm
\begin{abstract}
I illustrate a general formalism based upon the subtraction method 
for the calculation of next-to-leading order QCD cross 
sections for any number of jets in any type of hard collisions. 
I discuss the implementation of this formalism in a numerical 
program which generates partonic kinematical configurations with 
an appropriate weight, thus allowing the definition of arbitrary 
jet algorithms and cuts matching the experimental setup at the 
last step of the computation. I present results obtained with
computer codes which calculate one-jet and two-jet inclusive quantities
in photon-hadron and hadron-hadron collisions. 
\end{abstract}        
\vfill
\end{titlepage}
\section{Introduction}

The study of jet production provides one of the most fundamental
tests for the predictions of perturbative QCD. The jet rates are
rather large, and allow for some of the best precision measurements 
in hadron physics. One-jet and two-jet inclusive distributions
have been thoroughly measured in the past few years, and an impressive
amount of data for even more exclusive quantities, like three or
more jet cross sections, has been collected in $e^+e^-$ and 
hadron-hadron collisions. The high transverse momenta of the jets
set the scale for QCD calculations. The coupling constant is 
therefore small enough to result in a reliable 
perturbative series. Next-to-leading order predictions for one-jet 
and two-jet inclusive quantities in hadronic 
collisions~\mbox{[\ref{AGPG}-\ref{GGK}]} have been available for some
time. The uncertainty affecting these results is in general
smaller than the corresponding experimental errors, and a detailed 
comparison between theory and experiments has been carried out.
Although some issues need to be clarified (like, for example, the
tail of the $E_{\sss T}$ distribution measured by CDF~[\ref{CDFhighet}],
which appears to be higher than the theory), the overall agreement
is quite satisfactory. In the near future, the increased luminosity 
at HERA will also allow a statistically significant study of large
transverse momentum phenomena in photon-hadron collisions and DIS, 
thus giving another handle to test next-to-leading order QCD
predictions~[\ref{jetsatHERA}-\ref{jetsDIS}].

The calculation of jet cross sections at next-to-leading order is
rather complicated. A large number of infrared divergencies is found
in the computation of virtual and real diagram contributions,
due to the large number of colour-interacting, massless
partons involved in the hard scattering processes. It is then 
necessary to devise a procedure which allows the analytic calculation 
of the divergent parts and shows their cancellation in the sum which 
defines any infrared-safe physical observable. This task has been
accomplished in the past by using the slicing~[\ref{slicing}] and the 
subtraction methods. In ref.~[\ref{KS}] a formalism adopting the subtraction
method has been introduced in order to calculate one-jet and two-jet
inclusive quantities in hadronic collisions. Although possible, the extension 
of this formalism to other type of processes or more exclusive observables
is not straightforward. For this reason, a fully general formalism based 
on the subtraction method has been proposed in ref.~[\ref{FKS}],
where the results were reported in a form motivated by the study
of three-jet-like quantities in hadronic collisions. The formalism
has been subsequently applied to the production of four jets in $e^+e^-$ 
collisions in ref.~[\ref{SD}], and the same calculation techniques
have been used to study the hadroproduction of jets containing
heavy quarks in ref.~[\ref{FM}] (for other applications 
of the subtraction method, see refs.~[\ref{subtepem}-\ref{HS}]). 
Afterwards, a general formalism has been presented in ref.~[\ref{CS}], 
which uses the subtraction method to cancel the infrared divergencies and
a new method (called dipole method) to perform the analytic treatment
of the divergent terms. See also ref.~[\ref{NT}] for another approach.

The aim of this paper is to illustrate some further improvements of
the formalism of ref.~[\ref{FKS}], especially relevant for its implementation
in a computer code. I begin by writing the formulae of ref.~[\ref{FKS}]
in a form such that their generality is apparent, and the calculation
of any infrared-safe cross section in an arbitrary hard collision 
is straightforward. I then show how to construct a computer code 
(which I will call parton generator) that generates partonic events, 
which are eventually used to plot infrared-safe quantities. 
In this way, the jet reconstruction algorithm and the definition
of cuts matching the experimental conditions are inserted at the
last step of the computation. During the same computer run one
can therefore obtain cross sections for several, different, jet
definitions, as well as predictions for other infrared-safe quantities,
like for example shape variables. I point out that, in spite of
these features, such a parton generator is not equivalent to the usual
Monte Carlo parton shower programs, since it is the result of a 
fixed-order QCD calculation. Finally, I discuss the special case of 
processes with two or three partons in the final state, and I present 
few results obtained with parton generator codes\footnote{~The codes are 
available upon request.} written for photon-hadron and hadron-hadron 
collisions, suited for applications to HERA physics.

The paper is organized as follows: after presenting the formalism
in section 2, in section 3 I show how to write a parton generator code, 
and I present numerical results. I report my conclusions in section 4. 
Technical details concerning the formalism adopted are collected in 
appendix A and appendix B.
\section{Formalism}\label{sec:xsec}

The goal is to calculate the cross section for some infrared-safe
quantity in a given hard scattering process. To be specific, 
I start with the production of $N-1$ jets in hadronic collisions.
According to the factorization theorem in QCD~[\ref{factthref}], 
any differential cross section can be written as 
\beq
d\sigma^{(H_1 H_2)}(K_1,K_2)=\sum_{\aoat}\int dx_1 dx_2 f^{(H_1)}_{a_1}(x_1)
f^{(H_2)}_{a_2}(x_2)d\hat{\sigma}_{\aoat}(x_1 K_1,x_2 K_2)\,,
\label{factth}
\eeq
where $H_1$ and $H_2$ are the incoming hadrons, with momenta $K_1$ and
$K_2$ respectively, $f^{(H_i)}_{a_i}$ is the non-calculable but universal
distribution function for the parton $a_i$ in the hadron $H_i$,
and $d\hat{\sigma}_{\aoat}$ are the (subtracted) short-distance
partonic cross sections. As shown in ref.~[\ref{FKS}], the 
cancellation of the infrared divergencies\footnote{~In this paper, 
I will never deal with the problem of ultraviolet divergencies. I assume 
that they are renormalized in a proper way.} arising in the intermediate 
steps of the calculation at next-to-leading order can be fully performed at 
the level of the partonic cross sections. To prove this issue, the fact 
that initial state partons $a_1$ and $a_2$ are quarks and gluons is not
crucial. This implies that the same proof holds true for quantities
like $d\hat{\sigma}_{\gamma a_2}$ (entering photon-hadron cross sections),
$d\hat{\sigma}_{e a_2}$ (entering DIS cross sections), as well as for the
$e^+e^-$ cross section $d\hat{\sigma}_{e^+e^-}$. For this reason,
in the following I will only deal with partonic cross sections; it
is understood that initial state partons will have to be interpreted 
in a broad sense (that is, they can be quarks, gluons, photons and 
electrons, depending upon the type of physical hard scattering process 
one is interested to study. On the other hand, with final state partons
I will always mean quarks and gluons). We will get the $(N-1)$-jet 
cross section in the collision of particles $A$ and $B$ by using 
the equation
\beq
d\sigma^{(AB)}(K_1,K_2)=\sum_{\aoat}\int dx_1 dx_2 
L_{\aoat}^{(AB)}(x_1,x_2) d\hat{\sigma}_{\aoat}(x_1 K_1,x_2 K_2)\,,
\label{genfactth}
\eeq
where $L_{\aoat}^{(AB)}(x_1,x_2)$ is a suitable luminosity function. 
Eq.~(\ref{genfactth}) reduces to eq.~(\ref{factth}) with 
\mbox{$L_{\aoat}^{(H_1H_2)}(x_1,x_2)=
f^{(H_1)}_{a_1}(x_1)f^{(H_2)}_{a_2}(x_2)$}.
In the very same way, from eq.~(\ref{genfactth}) we get the photon-hadron 
cross section if we put
\beq
L_{\aoat}^{(\gamma H_2)}(x_1,x_2)=\delta_{\gamma a_1}\delta(1-x_1)
f^{(H_2)}_{a_2}(x_2), 
\eeq
the electron-hadron cross section in the Weizs\"acker-Williams
approximation with
\beq
L_{\aoat}^{(e H_2)}(x_1,x_2)=\delta_{\gamma a_1}f^{(e)}_{\gamma}(x_1) 
f^{(H_2)}_{a_2}(x_2)
\eeq
($f^{(e)}_{\gamma}$ is the Weizs\"acker-Williams function), the DIS 
cross section with
\beq
L_{\aoat}^{(e H_2)}(x_1,x_2)=\delta_{e a_1}\delta(1-x_1)
f^{(H_2)}_{a_2}(x_2), 
\eeq
and the $e^+e^-$ cross section with
\beq
L_{\aoat}^{(e^+e^-)}(x_1,x_2)=\delta_{e^+a_1}\delta(1-x_1)
\delta_{e^-a_2}\delta(1-x_2).
\eeq

At the next-to-leading order in QCD, I write the partonic cross sections
for the production of $(N-1)$-jets as 
\beq
d\hat{\sigma}_{\aoat}=d\hat{\sigma}_{\aoat}^{(0)}
+d\hat{\sigma}_{\aoat}^{(1)}\,.
\label{decomposition}
\eeq
The leading order term $d\hat{\sigma}_{\aoat}^{(0)}$ gets contributions 
from the processes
\beq
a_1(k_1)+a_2(k_2)\;\;\longrightarrow\;\;
\overbrace{a_3(k_3)+\cdots +a_{N+1}(k_{N+1})}^{N-1}
\;\;\Big(\,+\,e(k_e)\Big)\,,
\label{Nmoproc}
\eeq
where the final state partons $a_i$, $i=3,...,N+1$, are quarks and gluons, 
the electron in the final state is present only in the case of DIS
(in which case, $a_1=e$, $a_2=g,q,\bar{q}$). I will denote the processes of 
eq.~(\ref{Nmoproc}) as $(N-1)$-parton processes, apart from the presence 
of the electron in the final state. We can write
\beq
d\hat{\sigma}_{\aoat}^{(0)}=\uoNmofct\FLNmosum\MNmo(\FLNmofullj)
{\cal S}_{N-1}d\phi_{N-1}.
\label{borndef}
\eeq
In this equation, $\MNmo(\FLNmofullj)$ is the invariant amplitude for the 
process of eq.~(\ref{Nmoproc}), squared, summed over final state and averaged 
over initial state colour and spin degrees of freedom and multiplied by the 
flux factor, and $d\phi_{N-1}$ is the phase-space for $N-1$ massless 
partons (plus the electron in DIS). In order to include the contribution 
of all the partonic processes initiated by $a_1+a_2$, a sum over the 
flavours $g,u,\bar{u},...$ of final state partons has been performed; 
the statistical factor \mbox{$1/(N-1)!$} has therefore to be inserted to 
avoid double counting. The quantity ${\cal S}_{N-1}$, called measurement 
function, embeds the definition of the momenta of the $N-1$ jets
in terms of the momenta of the $N-1$ partons. I will discuss its properties
at length in the following.

The next-to-leading order term is
\beq
d\hat{\sigma}_{\aoat}^{(1)}=d\sigma_{\aoat}^{(v)}
+d\sigma_{\aoat}^{(r)}+d\sigma_{\aoat}^{(c)}\,,
\label{NLOdef}
\eeq
where
\beq
d\sigma_{\aoat}^{(v)}=\uoNmofct\FLNmosum\MNmoV(\FLNmofullj)
{\cal S}_{N-1}d\phi_{N-1}
\label{virtdef}
\eeq
is the contribution of the QCD loop corrections to processes 
in eq.~(\ref{Nmoproc}),
\beq
d\sigma_{\aoat}^{(r)}=\uoNfct\FLNsum\MN(\FLNfullj)
{\cal S}_N d\phi_N
\label{realdef}
\eeq
is the contribution of the tree amplitude of $N$-parton processes,
and $d\sigma_{\aoat}^{(c)}$ is the contribution of the initial 
state collinear counterterms (if needed). Eq.~(\ref{realdef})
is analogous to eq.~(\ref{borndef}). The $N-1$ jets have now to be
defined in terms of the $N$-body partonic kinematics: this is accomplished
by the measurement function ${\cal S}_N$.

Although the quantities defined in eqs.~(\ref{virtdef}) and~(\ref{realdef})
and the collinear counterterms are infrared divergent, their sum in 
eq.~(\ref{NLOdef}) is finite provided that the corresponding observable 
is infrared safe. As it was shown in refs.~[\ref{KS},\ref{FKS}], 
this requirement can be easily expressed in terms of the measurement 
functions. Explicitly, the conditions (infrared limits)
\beqn
&&\lim_{k_i^0\to 0} {\cal S}_N = {\cal S}_{N-1}\,,\;\;\;\;\;
\lim_{\vec{k}_i\parallel \vec{k}_j} {\cal S}_N = {\cal S}_{N-1}\,,
\label{IRsafe1}
\\&&
\lim_{\vec{k}_i\parallel \vec{k}_1} {\cal S}_N = {\cal S}_{N-1}\,,\;\;\;\;\;
\lim_{\vec{k}_i\parallel \vec{k}_2} {\cal S}_N = {\cal S}_{N-1}\,,
\label{IRsafe2}
\eeqn
with $3\leq i\leq N+2$, $3\leq j\leq N+2$, $i\neq j$, guarantee that 
the next-to-leading order contribution $d\hat{\sigma}_{\aoat}^{(1)}$ is 
finite. In eqs.~(\ref{IRsafe1}) and~(\ref{IRsafe2}), ${\cal S}_{N-1}$ 
is constructed  with the $(N-1)$-parton kinematics obtained from
the $N$-parton kinematics in the limits indicated.
I point out that the proof of ref.~[\ref{FKS}] has been carried out 
for $N=4$ (which corresponds to three-jet production); nevertheless, 
the fact that $N=4$ has never been explicitly used,
and therefore the proof is valid for an arbitrary $N$. Also, the proof
of ref.~[\ref{FKS}] exploits eqs.~(\ref{IRsafe1}) and~(\ref{IRsafe2}),
which in that case follow from the specific jet definition adopted.
Here I do not fix a jet definition, and therefore eqs.~(\ref{IRsafe1}) 
and~(\ref{IRsafe2}) play the r\^{o}le of conditions which {\it must}
be fulfilled by the jet definition chosen to induce an 
infrared-safe cross section. Finally, notice that eqs.~(\ref{IRsafe1}) 
and~(\ref{IRsafe2}) imply that most of the multiple infrared limits
of ${\cal S}_N$ vanish (by definition, in these limits two or more 
infrared divergencies overlap). This is because in these limits the 
$N$-parton kinematics reduces to a configuration where only $N-m$ 
partons ($m\ge 2$) have non-vanishing transverse momentum, and it is 
not possible to define $N-1$ jets with less than $N-1$ hard transverse 
partons. The only non-vanishing multiple infrared limits are the 
soft-collinear ones
\beq
\lim_{k_i^0\to 0,\vec{k}_i\parallel \vec{k}_j} {\cal S}_N = {\cal S}_{N-1}
\,,\;\;\;
\lim_{k_i^0\to 0,\vec{k}_i\parallel \vec{k}_1} {\cal S}_N = {\cal S}_{N-1}
\,,\;\;\;
\lim_{k_i^0\to 0,\vec{k}_i\parallel \vec{k}_2} {\cal S}_N = {\cal S}_{N-1}
\,.
\label{Ssoftcolllim}
\eeq
I remark that multiple infrared  configurations which are not 
soft-collinear can contribute to QCD cross sections beyond 
next-to-leading order.

In order to prove that eq.~(\ref{NLOdef}) is finite, one has to evaluate
analytically, with some suitable regularization, eq.~(\ref{virtdef}), 
eq.~(\ref{realdef}) and the collinear counterterms and then show that 
the divergent terms mutually cancel. The structure of the divergencies
in eq.~(\ref{virtdef}) naturally arises from the calculation of
loop integrals. The case of eq.~(\ref{realdef}) is more involved:
in fact, the divergencies are due to the integration over the 
regions of the phase space where one parton is soft, or two partons 
are collinear (which I will call infrared singular regions).
Due to the complexity of the $N$-body kinematics, one can 
not perform the analytic integration over the whole phase
space. The best one can do is to deal with one soft-collinear
singularity at a time. To achieve this goal in the framework
of the subtraction method, in ref.~[\ref{KS}] the $N$-body matrix
elements squared (in that case, $N=3$) were decomposed into single-singular
terms (having, by definition, one soft-collinear singularity at most); 
each term was then integrated over the relevant infrared singular
region. Although in principle this method can be extended to 
larger values of $N$, the amount of algebraic calculations and
analytic integrations required grows very rapidly, and poses
serious difficulties already with $N=4$. In ref.~[\ref{FKS}] a
different approach was proposed to overcome this problem. The key
idea is to use the properties of the measurement functions
to integrate over the infrared singular regions. In particular,
the following decomposition can be exploited
\beq
{\cal S}_N=\sum_{i=3}^{N+2}\left({\cal S}_i^{(0)} 
+ \sum_{\stackrel{j=3}{j\neq i}}^{N+2} {\cal S}_{ij}^{(1)} 
\theta(k_{j{\sss T}}^2-k_{i{\sss T}}^2)\right).
\label{Sfundecomp}
\eeq
The terms in the RHS of this equation are defined by their behaviour 
close to the infrared singular regions. In particular
\beqn
&&{\cal S}_i^{(0)}\neq 0\;\;\;\; {\rm only~if}\;\;\;\;
k_i^0\,\to\,0,\;\;\vec{k}_i\parallel\vec{k}_1\,,
\;\;\vec{k}_i\parallel\vec{k}_2\,,
\label{S0sing}
\\&&
{\cal S}_{ij}^{(1)}\neq 0\;\;\;\; {\rm only~if}\;\;\;\;
k_i^0\,\to\,0,\;\;k_j^0\,\to\,0,\;\;
\vec{k}_i\parallel\vec{k}_j\,.
\label{S1sing}
\eeqn
I stress that ${\cal S}_i^{(0)}$ and ${\cal S}_{ij}^{(1)}$ vanish in the 
infrared limits not explicitly indicated in eqs.~(\ref{S0sing}) 
and~(\ref{S1sing}). Two remarks are in order here. Firstly, 
eqs.~(\ref{S0sing}) and~(\ref{S1sing}) only constrain the infrared limits of 
${\cal S}_i^{(0)}$ and ${\cal S}_{ij}^{(1)}$. Therefore, these quantities
can be redefined up to terms which vanish in these limits.
This is the case of the functions ${\cal S}_i^{(fin)}$ introduced in 
ref.~[\ref{FKS}], which have been re-absorbed in the present paper into
${\cal S}_i^{(0)}$ and ${\cal S}_{ij}^{(1)}$ (I will show in 
appendix A that this can be consistently accomplished). 
Secondly, in ref.~[\ref{FKS}],
eqs.~(\ref{Sfundecomp}), (\ref{S0sing}) and~(\ref{S1sing}) have been
derived using a given jet definition. Nevertheless, as already observed
there, it is easy to understand that ${\cal S}_i^{(0)}$ and 
${\cal S}_{ij}^{(1)}$ fulfilling these equations can be defined starting 
from any infrared-safe prescription, since they are directly induced 
by eqs.~(\ref{IRsafe1}) and~(\ref{IRsafe2}). I will give explicit
examples in the following.

Inserting eq.~(\ref{Sfundecomp}) into eq.~(\ref{realdef}), and exploiting
eqs.~(\ref{S0sing}) and~(\ref{S1sing}), we see that 
\mbox{$d\sigma_{\aoat}^{(r)}$} is split into a sum of terms each 
of which has one soft-collinear singularity at most. Therefore,
this procedure is equivalent to a single-singular decomposition of
the matrix elements squared, without requiring any algebraic
computation. In the end, following ref.~[\ref{FKS}], the result for the 
next-to-leading order term is arranged as the sum of a $N$-parton 
contribution and of a $(N-1)$-parton contribution
\beq
d\hat{\sigma}_{\aoat}^{(1)}=d\hat{\sigma}_{\aoat}^{(1,N)}
+d\hat{\sigma}_{\aoat}^{(1,N-1)}.
\label{NLOres}
\eeq
I report in appendix A and appendix B the explicit form of the
quantities in the RHS of eq.~(\ref{NLOres}). Each term contributing
to eq.~(\ref{NLOres}) is finite, and therefore we have an operational
prescription for the numerical evaluation of an arbitrary jet cross section
in the framework of the subtraction method.
\section{Numerical calculations}

I now turn to the problem of implementing the formalism of the previous 
section in a computer code. As a benchmark example, I use two-jet production
in hadronic collisions, and define the jets through the algorithm 
introduced by Ellis and Soper in ref.~[\ref{EllSop}] and formulated 
in terms of the quantities
\beqn
d_i&=&k_{i{\sss T}}^2\,,
\label{ESdef1}
\\
R_{ij}^2&=&(\eta_i-\eta_j)^2+(\varphi_i-\varphi_j)^2\,,
\label{ESdef2}
\\
d_{ij}&=&min(k_{i{\sss T}}^2,k_{j{\sss T}}^2)\frac{R_{ij}^2}{D^2}\,.
\label{ESdef3}
\eeqn
The constant $D$ is the jet-resolution parameter. We have ($p_{\sss J_i}$
are the jet momenta)
\beqn
{\cal S}_i^{(0)}&=&\overline{\sum_{\sigma(J)}}
\delta\left(\vec{p}_{\sss J_1}-\vec{k}_j\right)
\delta\left(\vec{p}_{\sss J_2}-\vec{k}_l\right)
\nonumber \\*&&\times
\theta(min([d_i])-d_i)\theta\left(p_{\sss J_1T}-p_{1\sss T}^{min}\right)
\theta\left(p_{\sss J_2T}-p_{2\sss T}^{min}\right),
\label{S0ES}
\\
{\cal S}_{ij}^{(1)}&=&\overline{\sum_{\sigma(J)}}
\delta\left(\vec{p}_{\sss J_1}-\vec{k}_i-\vec{k}_j\right)
\delta\left(\vec{p}_{\sss J_2}-\vec{k}_l\right)
\nonumber \\*&&\times
\theta(min([d_{ij}])-d_{ij})\theta\left(p_{\sss J_1T}-p_{1\sss T}^{min}\right)
\theta\left(p_{\sss J_2T}-p_{2\sss T}^{min}\right),
\label{S1ES}
\eeqn
and ${\cal S}_3$ is defined through eq.~(\ref{Sfundecomp}).
Here, $\{i,j,l\}=\{3,4,5\}$; $\overline{\Sigma}_{\sigma(J)}$ denotes
the sum over the permutations of jet labels, with a normalization 
factor $1/2$ inserted. $min([d_i])$ ($min([d_{ij}])$) is the minimum 
of the quantities $d_\alpha$, $d_{\alpha\beta}$ with $d_i$ ($d_{ij}$)
excluded. $p_{1\sss T}^{min}$ and $p_{2\sss T}^{min}$ are the minimum 
observable transverse energies of the two jets (they are fixed, input 
parameters). Finally, when two partons are merged into a jet, 
eq.~(\ref{S1ES}), the jet three-momentum is defined as the sum of the 
parton three-momenta; other definitions would only imply changing the 
argument of the first $\delta$ in eq.~(\ref{S1ES}). When only two partons 
are present in the final state, we have
\beq
{\cal S}_2=\overline{\sum_{\sigma(J)}}
\delta\left(\vec{p}_{\sss J_1}-\vec{k}_3\right)
\delta\left(\vec{p}_{\sss J_2}-\vec{k}_4\right)
\theta\left(p_{\sss J_1T}-p_{1\sss T}^{min}\right)
\theta\left(p_{\sss J_2T}-p_{2\sss T}^{min}\right).
\label{S2ES}
\eeq
It is very easy to check explicitly that eqs.~(\ref{S0ES}), (\ref{S1ES})
and~(\ref{S2ES}) fulfill the conditions on the measurement functions
discussed in the previous section.

It is now possible to evaluate the quantity
\beq
<H>=\sum_{\aoat}\int H\,L_{\aoat}\left(d\hat{\sigma}_{\aoat}^{(0)}
+d\hat{\sigma}_{\aoat}^{(1)}\right),
\label{Haverage}
\eeq
where $L_{\aoat}$ is the relevant parton luminosity 
and $H$ is any function of the jet momenta. We may think of $H$ as 
a product of $\theta$ functions, implementing experimental cuts and
selecting a bin of a given histogram. $<H>$ will therefore be interpreted
as the next-to-leading order QCD prediction for the cross section in 
that bin. The rules for the numerical evaluation of eq.~(\ref{Haverage}) 
can be read from eqs.~(\ref{borndef}), (\ref{sigiinfin}), (\ref{sigoutijfin}),
(\ref{sigNmoV}) and~(\ref{sigNmoR}). Schematically, one has to
perform the following operations:
\begin{itemize}
\begin{enumerate}
\item generate the Bjorken $x$'s and the partonic kinematics;
\item evaluate the weight and the jet momenta for this kinematical 
configuration, as specified by eq.~(\ref{Haverage}) and 
by the expressions of the partonic cross 
sections in eqs.~(\ref{borndef}), (\ref{sigiinfin}), (\ref{sigoutijfin}), 
(\ref{sigNmoV}) and~(\ref{sigNmoR}). Call an output routine with the 
weight and the jet momenta as entries;
\item in the output routine, put the weight in the histogram bin 
selected by the jet momenta.
\end{enumerate}
\end{itemize}
I stress that the real computation is actually more complicated
than the one I outlined above, since eqs.~(\ref{sigiinfin}), 
(\ref{sigoutijfin}) and~(\ref{sigNmoR}) require subtractions.
I will not discuss here the Monte Carlo calculation of a 
subtracted quantity, which is by now a standard procedure. 
The interested reader can find a thorough discussion
in ref.~[\ref{MNR}]. Following the prescriptions implicit in 
eq.~(\ref{Haverage}), it is possible to write a computer code
which can calculate $<H>$ for any well-defined quantity $H$.
I call this code a {\it jet generator}. For a given choice of input
parameters, it returns event by event the jet momenta (as defined by
the measurement functions), which are eventually put by the user 
in some histogram bin.

By using the formulae collected in appendix~A and~B, the construction
of a jet generator is therefore fairly straightforward, and allows the
computation of the cross section for the production of any number of jets 
in any hard collision in the framework of the subtraction method, without
requiring algebraic manipulations of the partonic transition amplitudes,
as in ref.~[\ref{KS}]. The measurement functions, embedding the jet 
definition, can be written by the user in his own computer routine. 

The main drawback of such a jet generator is the following: the jet
definition is used, through the measurement functions, to disentangle
the singularities appearing in the real contribution. Therefore, to
get the jet cross section for several, different, jet definitions,
one has to perform several computer runs. Furthermore, a jet generator
outputs only jet momenta, which implies that it is not possible to
calculate non-jet-like infrared-safe observables, as for example 
shape variables. I will now show that it is not difficult to overcome
these problems using the formalism of ref.~[\ref{FKS}]. To be specific,
I start by considering again two-jet production is hadronic collisions.
With three partons in the final state, I introduce the following 
quantities, defined in terms of the partonic momenta
\beqn
{\cal P}_i^{(0)}&=&\Theta_i^{(0)}\,\theta(k_{3{\sss T}}
+k_{4{\sss T}}+k_{5{\sss T}}-E_{\sss T}^{min}),
\label{P0def}
\\
{\cal P}_{ij}^{(1)}&=&\Theta_{ij}^{(1)}\,\theta(k_{3{\sss T}}
+k_{4{\sss T}}+k_{5{\sss T}}-E_{\sss T}^{min}),
\label{P1def}
\\
{\cal P}_3&=&\sum_{i=3}^5\left({\cal P}_i^{(0)}
+\sum_{\stackrel{j=3}{j\neq i}}^5 {\cal P}_{ij}^{(1)}
\theta(k_{j{\sss T}}^2-k_{i{\sss T}}^2)\right),
\label{P3part}
\eeqn
where $\Theta_i^{(0)}$ and $\Theta_{ij}^{(1)}$ are suitable products
of $\theta$ functions. With two partons in the final state, I define
\beq
{\cal P}_2=1\cdot\theta(k_{3{\sss T}}+k_{4{\sss T}}-E_{\sss T}^{min}).
\label{P2part}
\eeq
I require that, close to the infrared singular regions, the quantities
defined in eqs.~(\ref{P0def}) and~(\ref{P1def}) have the following
properties
\beqn
&&{\cal P}_i^{(0)}\neq 0\;\;\;\; {\rm only~if}\;\;\;\;
k_i^0\,\to\,0,\;\;\vec{k}_i\parallel\vec{k}_1,
\;\;\vec{k}_i\parallel\vec{k}_2\,,
\label{P0sing}
\\&&
{\cal P}_{ij}^{(1)}\neq 0\;\;\;\; {\rm only~if}\;\;\;\;
k_i^0\,\to\,0,\;\;k_j^0\,\to\,0,\;\;
\vec{k}_i\parallel\vec{k}_j
\label{P1sing}
\eeqn
(which means that they vanish in the infrared limits not explicitly 
indicated), and that they are such that (a relabeling of the partons is 
understood, if necessary)
\beqn
&&\lim_{k_i^0\to 0} {\cal P}_3 = {\cal P}_2\,,\;\;\;\;\;
\lim_{\vec{k}_i\parallel \vec{k}_j} {\cal P}_3 = {\cal P}_2\,,
\label{IRsafe1part}
\\&&
\lim_{\vec{k}_i\parallel \vec{k}_1} {\cal P}_3 = {\cal P}_2\,,\;\;\;\;\;
\lim_{\vec{k}_i\parallel \vec{k}_2} {\cal P}_3 = {\cal P}_2\,.
\label{IRsafe2part}
\eeqn
Furthermore, the following equation must be fulfilled
\beq
\sum_{i=3}^5\left(\Theta_i^{(0)}+\sum_{\stackrel{j=3}{j\neq i}}^5 
\Theta_{ij}^{(1)}\theta(k_{j{\sss T}}^2-k_{i{\sss T}}^2)\right)\equiv 1.
\label{covering}
\eeq
If we choose the free parameter $E_{\sss T}^{min}$ such that
\mbox{$E_{\sss T}^{min} < 2\,min\left(p_{1\sss T}^{min},
p_{2\sss T}^{min}\right)$}, eq.~(\ref{covering}) implies that
\beq
{\cal S}_3\equiv {\cal S}_3 {\cal P}_3\,,\;\;\;\;\;\;
{\cal S}_2\equiv {\cal S}_2 {\cal P}_2\,,
\label{SvsP}
\eeq
where the ${\cal S}$ functions were defined at the beginning of this
section (notice that it is always possible to choose $E_{\sss T}^{min}$ 
without any reference to a specific jet definition. For example, 
$E_{\sss T}^{min}$ can be less than twice the minimum transverse 
momentum observable by the detector). I now define quantities 
analogous to the partonic cross sections appearing in eqs.~(\ref{borndef}) 
and~(\ref{NLOres}) (see also appendix A), by formally substituting 
the ${\cal S}$ functions with the ${\cal P}$ functions. I adopt 
the following notation
\beqn
d\hat{\sigma}_{\aoat}^{(0)}({\cal P}_2)&=&
d\hat{\sigma}_{\aoat}^{(0)}\mid_{{\cal S}_2\to {\cal P}_2}\,,
\label{listP1}
\\
d\hat{\sigma}_{\aoat}^{(1,2)}({\cal P}_2)&=&
d\hat{\sigma}_{\aoat}^{(1,2)}\mid_{{\cal S}_2\to {\cal P}_2}\,,
\\
d\sigma_{\aoat,i}^{(in,f)}({\cal P}_i^{(0)})&=&
d\sigma_{\aoat,i}^{(in,f)}\mid_{{\cal S}_i^{(0)}\to {\cal P}_i^{(0)}}\,,
\\
d\sigma_{\aoat,ij}^{(out,f)}({\cal P}_{ij}^{(1)})&=&
d\sigma_{\aoat,ij}^{(out,f)}\mid_{{\cal S}_{ij}^{(1)}\to
{\cal P}_{ij}^{(1)}}\,,
\\
d\hat{\sigma}_{\aoat}^{(1,3)}({\cal P}_3)&=&\sum_{i=3}^{5}\left(
d\sigma_{\aoat,i}^{(in,f)}({\cal P}_i^{(0)})
+\sum_{\stackrel{j=3}{j\neq i}}^{5} 
d\sigma_{\aoat,ij}^{(out,f)}({\cal P}_{ij}^{(1)})\right).
\label{listP5}
\eeqn
The quantity 
\beq
[H {\cal S}]=\sum_{\aoat}\int \,L_{\aoat}\left(
H {\cal S}_2\, d\hat{\sigma}_{\aoat}^{(0)}({\cal P}_2)
+ H {\cal S}_2\, d\hat{\sigma}_{\aoat}^{(1,2)}({\cal P}_2)
+ H {\cal S}_3\, d\hat{\sigma}_{\aoat}^{(1,3)}({\cal P}_3)\right)
\label{partgen}
\eeq
can be numerically evaluated, being finite. This can be easily
understood by observing that eqs.~(\ref{listP1})-(\ref{listP5})
are finite thanks to eqs.~(\ref{P0sing})-(\ref{IRsafe2part});
the proof is identical to the proof of ref.~[\ref{FKS}], which
showed that eqs.~(\ref{IRsafe1}), (\ref{IRsafe2}), (\ref{S0sing})
and~(\ref{S1sing}) guarantee the finiteness of the next-to-leading 
order cross section. Furthermore, thanks to eq.~(\ref{SvsP}), we have
\beq
<H>=[H {\cal S}]\,,
\eeq
where $<H>$ was evaluated in eq.~(\ref{Haverage}). Therefore, every
quantity which can be calculated with eq.~(\ref{Haverage}) can
also be calculated with eq.~(\ref{partgen}). Nevertheless,
eq.~(\ref{partgen}) is much more flexible. In fact, its numerical 
evaluation requires the following steps
\begin{itemize}
\begin{enumerate}
\item generate the Bjorken $x$'s and the partonic kinematics;
\item evaluate the weight for this kinematical configuration, 
as specified by eq.~(\ref{partgen}) and by the expressions of the 
partonic cross sections in eqs.~(\ref{listP1})-(\ref{listP5}). 
Call an output routine with the weight and the parton momenta as entries;
\item in the output routine, define the jet momenta as specified by the
jet-finding algorithm embedded in the ${\cal S}$ functions, and put the 
weight in the histogram bin selected by the jet momenta.
\end{enumerate}
\end{itemize}
Therefore, the code implementing eq.~(\ref{partgen}) returns event
by event the {\it parton} momenta, which are eventually manipulated by
the user. I call such a code a {\it parton generator}.
The main difference between a jet generator and a parton generator
can be read from eqs.~(\ref{Haverage}) and~(\ref{partgen}); while
in eq.~(\ref{Haverage}) the measurement functions ${\cal S}$ enter
the partonic cross sections and render them finite, in eq.~(\ref{partgen})
they have the same r\^{o}le of the function $H$. In eq.~(\ref{partgen})
the partonic cross sections are finite thanks to the ${\cal P}$ functions, 
which can be chosen once and forever without any reference to a specific
jet definition. For this reason, such a parton generator is able to plot, 
on an event-by-event basis, any infrared-safe quantity defined with two or 
three partons in the final state. Namely, in a single run it can produce 
one-jet and two-jet observables, with the jets defined by several
algorithms, as well as various shape variables. Notice that
eqs.~(\ref{P2part}) and~(\ref{covering}) guarantee that the events are 
generated in the whole phase space. Some of them are eventually rejected, 
namely those having a kinematics which is not fulfilling
\beqn
&&\theta(k_{3{\sss T}}+k_{4{\sss T}}+k_{5{\sss T}}-E_{\sss T}^{min}),
\label{ETcut1}
\\
&&\theta(k_{3{\sss T}}+k_{4{\sss T}}-E_{\sss T}^{min}).
\label{ETcut2}
\eeqn
Nevertheless, $E_{\sss T}^{min}$ can always be chosen in such a way
that eqs.~(\ref{ETcut1}) and~(\ref{ETcut2}) are less stringent than 
the physical cuts which are required to define any infrared-safe quantity. 
Technically, eqs.~(\ref{ETcut1}) and~(\ref{ETcut2}) are inserted to avoid 
multiple non soft-collinear infrared singularities, which do not contribute 
to the cross section at next-to-leading order. In the soft-collinear
limits we have, analogously to eq.~(\ref{Ssoftcolllim}),
\beq
\lim_{k_i^0\to 0,\vec{k}_i\parallel \vec{k}_j} {\cal P}_3 = {\cal P}_2
\,,\;\;\;
\lim_{k_i^0\to 0,\vec{k}_i\parallel \vec{k}_1} {\cal P}_3 = {\cal P}_2
\,,\;\;\;
\lim_{k_i^0\to 0,\vec{k}_i\parallel \vec{k}_2} {\cal P}_3 = {\cal P}_2
\,.
\label{Psoftcolllim}
\eeq

It should be clear that the method described so far can be used,
without {\it any} modification, to write a parton generator for
photon-hadron or $e^+e^-$ collisions. In the latter case, we can also
set $E_{\sss T}^{min}=0$, since multiple soft singularities are forbidden
by energy conservation and configurations with partons collinear
to the incoming leptons are not singular.

The situation is slightly more complicated with more than three particles
in the final state in DIS (two or more jet production), and
for three or more jet production in hadron-hadron, photon-hadron and
$e^+e^-$ collisions. Indeed, it is easy to understand that with 
a four-body (or more) kinematics, the analogous of eqs.~(\ref{ETcut1})
and~(\ref{ETcut2}) are not enough to avoid multiple non soft-collinear
infrared singularities. Eq.~(\ref{ETcut1}) and eq.~(\ref{ETcut2}) have 
to be substituted by suitable products of $\theta$ functions which 
vanish in the multiple non soft-collinear infrared regions. 
In the end, one should get quantities ${\cal P}_N$ and
${\cal P}_{N-1}$, defined analogously to ${\cal P}_3$ and ${\cal P}_2$
in eqs.~(\ref{P3part}) and~(\ref{P2part}), which fulfill
eqs.~(\ref{P0sing})-(\ref{covering}) and~(\ref{Psoftcolllim}) (with
the formal substitutions $2\to N-1$, $3\to N$).
The products of $\theta$ functions which substitute eqs.~(\ref{ETcut1}) 
and~(\ref{ETcut2}) must be chosen in such a way that
\beq
{\cal S}_N\equiv {\cal S}_N {\cal P}_N\,,\;\;\;\;\;\;
{\cal S}_{N-1}\equiv {\cal S}_{N-1} {\cal P}_{N-1}\,.
\label{SvsPNpart}
\eeq
As already observed before, I point out that, in the case of hadron-hadron,
photon-hadron and $e^+e^-$ collisions with two or three partons in the
final state, with a single choice of the ${\cal P}$ functions it is
possible to fulfill eq.~(\ref{SvsPNpart}) for any measurement
functions ${\cal S}$. The same is not true for the other cases.
In practice, if one chooses the ${\cal P}$ functions in such a way
that they only vanish {\it extremely} close to the multiple non soft-collinear
infrared regions, then eq.~(\ref{SvsPNpart}) holds for all the physically
meaningful choices of the measurement functions ${\cal S}$.
We therefore get again a parton generator which is able to plot,
event by event, $(N-1)$-jet inclusive quantities with several
different jet definitions, and shape variables which get contributions
at next-to-leading order from $(N-1)$- and $N$-parton configurations.

In order to write a code for a parton generator, a definite choice
for the ${\cal P}$ functions has to be made. Although these functions 
can be rather freely chosen, the only constraints being 
eqs.~(\ref{P0sing})-(\ref{covering}) and~(\ref{Psoftcolllim}), 
their properties are motivated by the properties of the measurement 
functions of a jet algorithm. Therefore, it is quite natural to use a 
jet algorithm to construct the ${\cal P}$ functions. Notice that, by 
definition, this jet algorithm has nothing to do with the jet algorithm(s) 
eventually used to obtain predictions for physical observables.

I now consider the case of photon-hadron and hadron-hadron collisions
with two or three partons in the final state. Using the prescription of 
ref.~[\ref{EllSop}], we get, as can be also directly seen from 
eqs.~(\ref{S0ES}) and~(\ref{S1ES}),
\beqn
\Theta_i^{(0)}&=&\theta(min([d_i])-d_i),
\label{P0ES}
\\
\Theta_{ij}^{(1)}&=&\theta(min([d_{ij}])-d_{ij}),
\label{P1ES}
\eeqn
where the relevant quantities have been introduced in 
eqs.~(\ref{ESdef1})-(\ref{ESdef3}). In order to fulfill eq.~(\ref{covering}),
we must have $D<2\pi/3$. If we adopt the prescription of the cone
algorithm~[\ref{conealg}], we get
\beqn
\Theta_i^{(0)}&=&\theta(R_{ij}-g_{ij})\theta(R_{il}-g_{il})
\theta(R_{jl}-g_{jl})\theta(min(k_{j{\sss T}},k_{l{\sss T}})-k_{i{\sss T}}),
\label{P0cone}
\\
\Theta_{ij}^{(1)}&=&\theta(g_{ij}-R_{ij})\theta(R_{il}-g_{il})
\theta(R_{jl}-g_{jl})\theta(k_{l{\sss T}}-min(k_{i{\sss T}},k_{j{\sss T}})),
\label{P1cone}
\eeqn
where
\beq
g_{ij}=\frac{k_{i{\sss T}}+k_{j{\sss T}}}
{max(k_{i{\sss T}},k_{j{\sss T}})} \,R.
\eeq
Eq.~(\ref{covering}) requires that $R<\pi/3$.

I wrote a fortran code for a parton generator in photon-hadron 
collisions, and a fortran code for a parton generator in hadron-hadron
collisions. The latter can be used also in the case of hadronic photon-hadron
collisions, that is for photon-hadron interactions where the photon 
fluctuates into a hadronic state before undergoing a hard collision.
Therefore, the codes are suitable for applications to HERA physics.
In order to test the codes, I produced single-inclusive jet and
two-jet observables. The results obtained with the photon-hadron 
(hadron-hadron) code have been compared with the results of 
ref.~[\ref{KKdijet}] (refs.~[\ref{KKdijet},\ref{GGK}]). In both cases,
I found nice agreement.

\begin{figure}
\centerline{\epsfig{figure=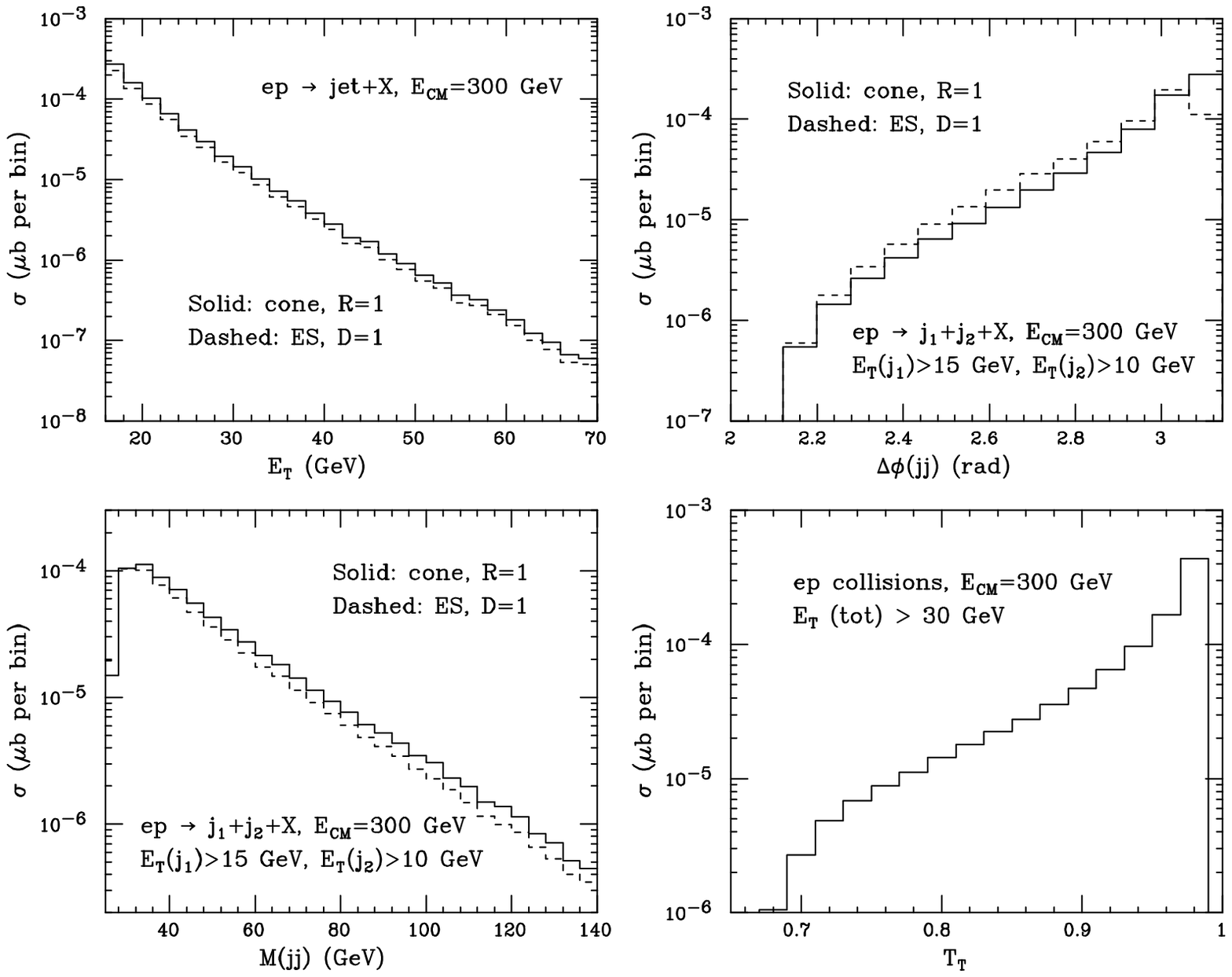,width=\textwidth,clip=}}
\ccaption{}{ \label{fig:pnt}  
Jet observables and transverse thrust in $ep$ collisions 
(Weizs\"acker-Williams approximation) at HERA. Pointlike photon only.}
\end{figure}

\begin{figure}
\centerline{\epsfig{figure=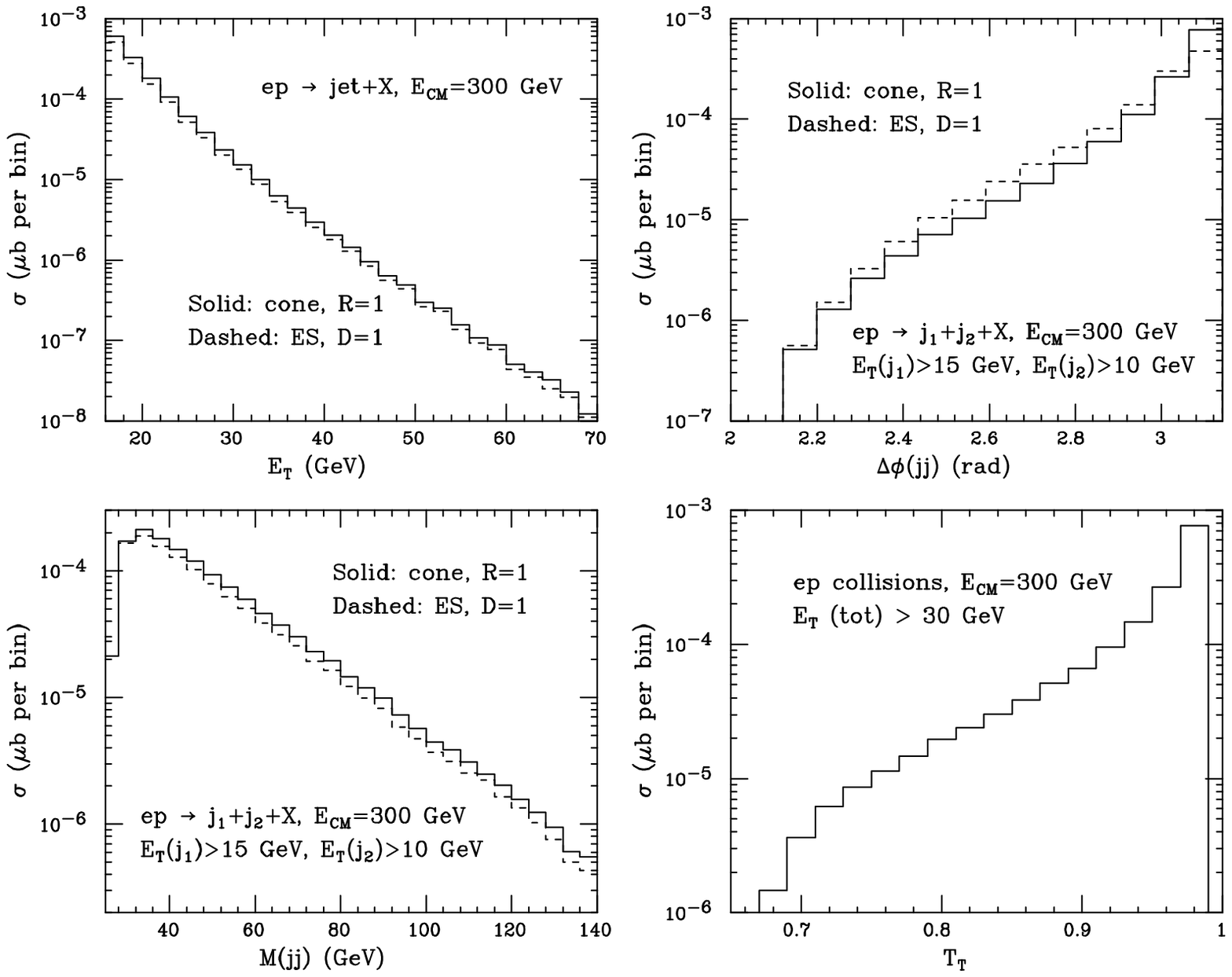,width=\textwidth,clip=}}
\ccaption{}{ \label{fig:had}  
Jet observables and transverse thrust in $ep$ collisions 
(Weizs\"acker-Williams approximation) at HERA. Hadronic photon only.}
\end{figure}

In figures~\ref{fig:pnt} and~\ref{fig:had} I show the transverse
energy of the single-inclusive jet, the azimuthal distance and the
invariant mass of the pair of the two hardest jets, and the
transverse thrust distributions for $ep$ collisions in 
the HERA energy range, $E_{\sss CM}=300$~GeV, as predicted by 
the aforementioned parton generators.
%%%%%%%%%%%%%%%%%%%%%%%%%%%%%%%%%%%%%%%%%%%%%%%%%%%%%%%%%%%%%%%%%%%%%%
%%%%%%%%%%%%%%%%%%%%%%%%%%%%%%%%%%%%%%%%%%%%%%%%%%%%%%%%%%%%%%%%%%%%%%
The Weizs\"acker-Williams approximation has been adopted (the form
of ref.~[\ref{WWFMNR}], which includes non-logarithmic singular terms,
has been used with $Q_{\sss WW}^2=4$~GeV$^2$ and $0.2<y<0.8$), and 
therefore we are dealing with a photoproduction process. As such, both 
the pointlike photon cross section (whose contribution, obtained with
the photon-hadron code, is shown in figure~\ref{fig:pnt})
and the hadronic photon cross section (figure~\ref{fig:had}, obtained
with the hadron-hadron code) are sizeable. I point out that the
curves presented in the figures are not physical quantities,
if taken separately, since only their sum is measurable. Nevertheless,
%%%%%%%%%%%%%%%%%%%%%%%%%%%%%%%%%%%%%%%%%%%%%%%%%%%%%%%%%%%%%%%%%%%%%%
%%%%%%%%%%%%%%%%%%%%%%%%%%%%%%%%%%%%%%%%%%%%%%%%%%%%%%%%%%%%%%%%%%%%%%
here I am only interested in discussing the properties of parton
generator codes. Phenomenological results for HERA physics will
be reported in a forthcoming publication~[\ref{FR}]. The four plots 
appearing in the same figure have been obtained with a single computer
run. The jets have been defined using the cone algorithm~[\ref{conealg}]
with $R=1$ (solid histograms), and the algorithm of ref.~[\ref{EllSop}], 
with $D=1$ (dashed histograms). In the same run, I also computed the 
jet cross sections with the cone algorithm and $R=0.7$. I do not show the
results, since they are extremely close to the dashed histograms.
I used the set MRSA$^\prime$ for the parton densities 
in the proton, and the set GRV-HO for the parton densities
in the photon. The value of $\Lambda_{\sss QCD}$ was fixed at the
value suggested by the MRSA$^\prime$ set, $\Lambda_5=152$~MeV.
Factorization and renormalization scales have been taken equal to half of 
the total transverse energy of the event. The figures have been produced 
using eqs.~(\ref{P0cone}) and~(\ref{P1cone}), with $R=1$, to define
the ${\cal P}$ functions; I verified that completely equivalent
results can be obtained by using eqs.~(\ref{P0ES}) and~(\ref{P1ES})
with several values of $D$. This gives a consistency check on the
method used to construct the parton generator codes. Also notice
that the curves corresponding to the two different jet definitions 
adopted have comparable quality. Naively, one may think that the
curves relevant for the jet definition which matches the prescription
for the ${\cal P}$ functions could converge somewhat faster.
This is not true; the convergence properties are almost completely
dictated by the behaviour of the ${\cal P}$ functions and of the
jet definitions close to the infrared singular regions, where they
are strongly constrained by the infrared safeness conditions, which
are definition-independent and must hold in any case.

Finally, I would like to comment on the numerical calculations performed
with the subtraction method. The main drawback of the method is that
in the subtracted integrals the cancellation takes place between terms
which have different kinematics (see for example eq.~(\ref{uoxidef})).
For this reason, in ref.~[\ref{FKS}] the arbitrary parameters 
$\xi_{cut}$, $\delta_{\sss I}$, $\delta_o$ have been introduced in 
the calculation (see also refs.~[\ref{MNR}-\ref{subt}]). To be specific,
consider eq.~(\ref{uoxidef}); we see that the counterevent 
(that is, the integrand function calculated on the pole, $f(0)$) is 
subtracted only if \mbox{$\xi_i<\xi_{cut}$}. Therefore, with a suitable
choice of $\xi_{cut}$, the counterevent is subtracted only if the
event is close to the soft limit $\xi_i=0$. In this way, the
subtraction is performed only in those cases when the kinematics
of the event and of the counterevent are quite close to each other,
thus resulting in an improved numerical stability of the result.
As a bonus, we save computing time, since in most of the cases the
calculation of the counterevent is not necessary. It is very important
to realize that the parameters $\xi_{cut}$, $\delta_{\sss I}$, $\delta_o$
have nothing to do with the non-physical parameters which {\it must} be
introduced in the slicing method. In the present case, the parameters do not
have to be small, since no approximation was performed in the intermediate
steps of the calculation. Therefore, there is no need to prove that
the physical cross section is independent of $\xi_{cut}$, $\delta_{\sss I}$, 
$\delta_o$, since this is true by construction. We can exploit this
fact as a powerful test of the numerical implementation of the 
formalism, by {\it verifying} that the cross section is a constant
with respect to the choice of $\xi_{cut}$, $\delta_{\sss I}$, $\delta_o$.
By choosing $(\xi_{cut},\delta_{\sss I},\delta_o)$ close to $(0,0,0)$,
we also see that the slicing method can be obtained as a limit of the 
subtraction method (for this to be rigorously true, the kinematics
of the events close to the infrared limits must be set equal to the
kinematics of the corresponding counterevents. For example, this
amounts to set $f(\xi_i)=f(0)$ for $\xi_i<\xi_{cut}$ in eq.(\ref{uoxidef})). 
In this case, as it is apparent from eqs.~(\ref{sigiinfin}), 
(\ref{sigoutijfin}) and~(\ref{sigNmoV}), cancellations between large 
numbers take place in the sum which defines the physical cross section. 
\section{Conclusions}

In this paper I have studied the definition of infrared-safe
cross sections at next-to-leading order in QCD. The formalism
of ref.~[\ref{FKS}], which is based on the subtraction method,
has been written in a form which clearly shows its generality
and universality. The formulae can be applied to any hard 
scattering process, with an arbitrary number of final state
partons. I then turned to the problem of the numerical computation 
of infrared-safe cross sections. It has been shown how to write a computer 
code which, event by event, outputs the momenta of the final state 
partons. The momenta are eventually used in an analysis routine to 
define the physical observables. The special case of infrared-safe 
quantities defined with two or three partons in the final state has 
been studied. Two codes, one dealing with photon-hadron 
collisions and one dealing with hadron-hadron collisions, 
have been written. Sample numerical results have been presented
for the case of $ep$ collisions in the Weizs\"acker-Williams
approximation. This is relevant for applications to HERA physics,
where both the pointlike photon and the hadronic photon give
sizeable contributions to the physical cross section. I presented
predictions for one-jet and two-jet observables (as an example, I defined 
the jets using two different prescriptions), and for transverse thrust.

\section*{Acknowledgements}

It is a pleasure to thank Z.~Kunszt, P.~Nason and G.~Ridolfi
for useful discussions. Part of the results presented in this
paper are due to a collaboration with Z.~Kunszt and G.~Ridolfi.
\newpage
\begin{reflist}

\item \label{AGPG}
 F.~Aversa, M.~Greco, P.~Chiappetta and J.~P.~Guillet, \zp{C46}{90}{253};\\
 \prl{65}{90}{401}.
\item \label{EKS}
 S.~D.~Ellis, Z.~Kunszt and D.~E.~Soper, \prl{64}{90}{2121};\\
 \prl{69}{92}{1496};\\
 S.~D.~Ellis and D.~E.~Soper, \prl{74}{95}{5182}.
\item \label{GGK}
 W.~T.~Giele, E.~W.~N.~Glover and D.~A.~Kosower, \prl{73}{94}{2019}.
\item \label{CDFhighet}
 CDF Coll., F.~Abe {\it et al.}, \prl{68}{92}{1104};\\
 \prl{70}{93}{1376}.
\item \label{jetsatHERA}
 L.~E.~Gordon and J.~K.~Storrow, \pl{B291}{92}{320};\\
 G. Kramer and S.~G.~Salesh, \zp{C61}{94}{277};\\
 D. B\"odeker, G. Kramer and S.~G.~Salesh, \zp{C63}{94}{471};\\
 M.~Klasen and G.~Kramer, \zp{C72}{96}{107}, \hepph{9511405};\\
 B.~W.~Harris and J.~F.~Owens, preprint FSU-HEP-970411, \hepph{9704324}. 
\item \label{KKdijet}
 M.~Klasen and G.~Kramer, preprint DESY 96-246, \hepph{9611450}.
\item \label{jetsDIS}
 E.~Mirkes and D.~Zeppenfeld, \pl{B380}{96}{205}, \hepph{9511448};\\
 S.~Catani and M.~H.~Seymour, proceedings of the Workshop {\it Future
 Physics at HERA}, p.~519, \hepph{9609521}. 
\item \label{slicing}
 W.~T.~Giele and E.~W.~N.~Glover, \pr{D46}{92}{1980};\\
 W.~T.~Giele, E.~W.~N.~Glover and D.~A.~Kosower, \np{B403}{93}{633}.
\item \label{KS}
 Z.~Kunszt and D.~E.~Soper, \pr{D46}{92}{192}.
\item\label{FKS}
 S.~Frixione, Z.~Kunszt and A.~Signer, \np{B467}{96}{399}, \hepph{9512328}.
\item \label{SD}
 A.~Signer and L.~Dixon, \prl{78}{97}{811}, \hepph{9609460};\\
 \hepph{9706285}.
\item \label{FM}
 S.~Frixione and M.~Mangano, \np{B483}{97}{321}, \hepph{9605270}.
\item \label{subtepem}
 R.~K.~Ellis, D.~A.~Ross and A.~E.~Terrano, \np{B178}{81}{421};\\
 Z.~Kunszt and P.~Nason, in {\it Z Physics at LEP 1}, eds. G.~Altarelli,
 R.~Kleiss and C.~Verzegnassi, Geneva, 1989.
\item \label{ZZ}
 B.~Mele, P.~Nason and G.~Ridolfi, \np{B357}{91}{409}.
\item \label{MNR}
 M.~Mangano, P.~Nason and G.~Ridolfi, \np{B373}{92}{295}.
\item \label{subt}
 S.~Frixione, P.~Nason and G.~Ridolfi, \np{B383}{92}{3};\\
 S.~Frixione, \np{B410}{93}{280};\\
 S.~Frixione, M.~Mangano, P.~Nason and G.~Ridolfi , \np{B412}{94}{225}.
\item \label{HS}
 B.~W.~Harris and J.~Smith, \np{B452}{95}{109}.
\item \label{CS}
 S.~Catani and M.~H.~Seymour, \pl{B378}{96}{287}, \hepph{9602277};\\
 \np{B485}{97}{291}, \hepph{9605323}.
\item \label{NT}
 Z.~Nagy and Z.~Trocsanyi, \np{B486}{97}{189}, \hepph{9610498}.
\item \label{factthref}
 J.~C.~Collins, D.~E.~Soper and G.~Sterman, in {\it Perturbative
 Quantum Chromodinamics}, 1989, ed. Mueller, World Scientific,
 Singapore, and references therein.
\item \label{EllSop}
 S.~D.~Ellis and D.~E.~Soper, \pr{D48}{93}{3160}.
\item \label{conealg}  
 F.~Aversa {\it et al.}, Proceedings of the Summer Study on 
 High Energy Physics, Research Directions for the Decade, 
 Snowmass, CO, Jun 25 - Jul 13, 1990.
\item \label{WWFMNR}
 S.~Frixione, M.~Mangano, P.~Nason and G.~Ridolfi, \pl{B319}{93}{339}.
\item \label{FR}
 S.~Frixione and G.~Ridolfi, in preparation.
\item \label{ES}
 R.~K.~Ellis and J.~Sexton, \np{B269}{86}{445}.
\item \label{phscheme}
 M.~Gl\"uck, E.~Reya and A.~Vogt, \pr{D45}{92}{3986};\\
 S.~Frixione, M.~Mangano, P.~Nason and G.~Ridolfi, \hepph{9702287},
 to appear in {\it Heavy Flavours II}, eds. A.~J.~Buras and M.~Lindner, 
 World Scientific.

\end{reflist}
\newpage
\appendix
\section{N-parton contribution}

The $N$-parton contribution, first term in the RHS of eq.~(\ref{NLOres}),
is written as the sum of finite terms, whose form is induced by the 
decomposition of the measurement function of eq.~(\ref{Sfundecomp})
\beq
d\hat{\sigma}_{\aoat}^{(1,N)}=\sum_{i=3}^{N+2}\left(
d\sigma_{\aoat,i}^{(in,f)}
+\sum_{\stackrel{j=3}{j\neq i}}^{N+2} d\sigma_{\aoat,ij}^{(out,f)}\right).
\label{Nbodyxsec}
\eeq
The first term in the RHS of this equation is proportional to
${\cal S}_i^{(0)}$, the second one to ${\cal S}_{ij}^{(1)}$.
By recalling the properties of the measurement functions close to
the infrared limits, it is easy to choose a set of variables
suitable to perform the calculation with the subtraction method.
In the partonic center-of-mass frame, we parametrize the momentum
of the final state parton $i$ as
\beq
k_i=\frac{\sqrt{S}}{2}\xi_i\left(1,\sqrt{1-y_i^2}
\vec{e}_{i{\sss T}},y_i\right),
\label{kidef}
\eeq
where $\sqrt{S}$ is the partonic center-of-mass energy. We get~[\ref{FKS}]
\beqn
d\sigma_{\aoat,i}^{(in,f)}&=&\uoNfct\frac{1}{2}\uoxiic
\Bigg[\uoyimdi+\uoyipdi\Bigg]
\frac{1}{2(2\pi)^{3}}\left(\frac{\sqrt{S}}{2}\right)^{2}
\nonumber \\*&\times&
\left((1-y_i^2)\xi_i^2\FLNsum\MN(\FLNfullj)\right) 
{\cal S}_i^{(0)} d\phi d\xi_i dy_i d\varphi_i\,.
\label{sigiinfin}
\eeqn
The measure $d\phi$ is implicitly defined by writing the phase space 
for $N$ partons (plus electron in DIS) in $4$ dimensions as
\beq
d\phi_N=d\phi\,\frac{1}{2(2\pi)^3}
\left(\frac{\sqrt{S}}{2}\right)^2\xi_i
d\xi_i dy_i d\varphi_i\,.
\label{dphiNin}
\eeq
The statistical factor $1/N!$ is due to the sum over the flavours of 
final state partons (see eq.(\ref{realdef})). The damping factor
$(1-y_i^2)\xi_i^2$ and ${\cal S}_i^{(0)}$ guarantee that the integrand 
is finite everywhere in the phase space. We can therefore use
\beqn
\int_0^1 d\xi_i f(\xi_i)\uoxiic &=& \int_0^1 d\xi_i
\frac{f(\xi_i)-f(0)\Th(\xi_{cut}-\xi_i)}{\xi_i}\,,
\label{uoxidef}
\\
\int_{-1}^{1}dy_i g(y_i)\uoyimpdi &=& \int_{-1}^{1}dy_i\,
\frac{g(y_i)-g(\pm 1)\Th(\pm y_i-1+\delta_{\sss I})}{1\mp y_i}\,,
\label{uoyipmdef}
\eeqn
which hold for any smooth functions $f$ and $g$. These prescriptions 
are almost identical to the usual $+$ prescription, except for
the fact that the value of the integrand function at the pole is subtracted
only if the integration variable satisfies the condition imposed by
the $\theta$ functions. The parameters
\beq
0<\xi_{cut}\leq 1\,,\;\;\;\;\;\; 0<\delta_{\sss I}\leq 2
\eeq
can be {\it arbitrarily} chosen in the indicated range. Finally, notice that
if ${\cal S}_i^{(0)}$ vanishes in the soft ($\xi_i\to 0$) and
collinear ($y_i\to \pm 1$) limits, eq.~(\ref{sigiinfin}) becomes
\beq
d\sigma_{\aoat,i}^{(in,f)}=\uoNfct\FLNsum\MN(\FLNfullj)
{\cal S}_i^{(0)} d\phi_N.
\label{siginlimit}
\eeq
This expressions is identical to the one induced by the functions
${\cal S}_i^{fin}$ in ref.~[\ref{FKS}], and therefore proves that 
the contribution of ${\cal S}_i^{fin}$ can be safely re-absorbed into
the contribution of ${\cal S}_i^{(0)}$.

In order to give the explicit expression for the second term in
the RHS of eq.~(\ref{Nbodyxsec}), we rewrite eq.~(\ref{kidef}) as
\beq
k_i=\frac{\sqrt{S}}{2}\xi_i\left(1,\hat{k}_i\right),\;\;
\hat{k}_i=\hat{p}_i R\,,\;\;\hat{p}_i=\left(\vec{0},1\right),
\eeq
where $R$ is a suitable matrix, and parametrize the momentum of 
parton $j$ as
\beq
k_j=\frac{\sqrt{S}}{2}\xi_j\left(1,\hat{k}_j\right),\;\;
\hat{k}_j=\hat{p}_j R\,,\;\;
\hat{p}_j=\left(\sqrt{1-y_j^2}\vec{e}_{j{\sss T}},y_j\right).
\label{kjdef}
\eeq
Therefore, in the limit $y_j\to 1$ the partons $i$ and $j$ become
collinear to each other. We get~[\ref{FKS}]
\beqn
d\sigma_{\aoat,ij}^{(out,f)}&=&\uoNfct\uoxiic\uoyjmdo
\left((1-y_j)\xi_i^2\xi_j\FLNsum\MN(\FLNfullj)\right)
\nonumber \\*&\times&
{\cal S}_{ij}^{(1)}\Th(k_{j{\sss T}}^2-k_{i{\sss T}}^2)
\left(\frac{1}{2(2\pi)^3}\left(\frac{\sqrt{S}}{2}\right)^2\right)^2
d\tilde{\phi}d\xi_i d\xi_j dy_i dy_j d\varphi_i d\varphi_j\,,
\label{sigoutijfin}
\eeqn
where 
\beq
\int_{-1}^{1}dy_j g(y_j)\uoyjmdo = \int_{-1}^{1}dy_j\,
\frac{g(y_j)-g(1)\Th(y_j-1+\delta_o)}{1-y_j}
\label{uoyjmdef}
\eeq
and $0<\delta_o\leq 2$. The phase-space for $N$ partons (plus electron 
in DIS) in $4$ dimensions has been written in the following way, thus
implicitly defining $d\tilde{\phi}$
\beq
d\phi_N=d\tilde{\phi}\,\left(\frac{1}{2(2\pi)^3}
\left(\frac{\sqrt{S}}{2}\right)^2\right)^2\xi_i \xi_j
d\xi_i d\xi_j dy_i dy_j d\varphi_i d\varphi_j\,.
\label{dphiNout}
\eeq
If ${\cal S}_{ij}^{(1)}$ vanishes in the soft ($\xi_i\to 0$) and 
collinear ($y_j\to 1$) limits, we get
\beq
d\sigma_{\aoat,ij}^{(out,f)}=\uoNfct\FLNsum\MN(\FLNfullj)
{\cal S}_{ij}^{(1)}\Th(k_{j{\sss T}}^2-k_{i{\sss T}}^2) d\phi_N.
\eeq
Like eq.~(\ref{siginlimit}), this shows that the contribution of
${\cal S}_i^{fin}$ can also be re-absorbed into the contribution 
of ${\cal S}_{ij}^{(1)}$.
\subsection{(N-1)-parton contribution}

We now turn to the second term in the RHS of eq.~(\ref{NLOres}). 
We write it in the following way
\beq
d\hat{\sigma}_{\aoat}^{(1,N-1)}=d\hat{\sigma}_{\aoat}^{(1,N-1v)}
+d\hat{\sigma}_{\aoat}^{(1,N-1r)}\,,
\label{Nmobodyxsec}
\eeq
where the first term reads~[\ref{FKS}]
\beqn
d\hat{\sigma}_{\aoat}^{(1,N-1v)}&=&
\frac{\as}{2\pi}\FLNmosum {\cal Q}(\FLNmofullj) d\sigma^{(0)}(\FLNmofullj)
\nonumber \\*&+&
\frac{\as}{2\pi}\,\frac{1}{2}\sum_{\stackrel{i,j=1}{i\neq j}}^{N+1}
{\cal I}_{ij}^{(reg)}\FLNmosum d\sigma_{ij}^{(0)}(\FLNmofullj)
\nonumber \\*&+&
\frac{\as}{2\pi}\uoNmofct\FLNmosum
\MNmoV_{\sss NS}(\FLNmofullj)\,{\cal S}_{N-1}\,d\phi_{N-1}\,.
\label{sigNmoV}
\eeqn
Here
\beqn
{\cal I}_{ij}^{(reg)}&=&\frac{1}{8\pi^2}\Bigg[
\frac{1}{2}\log^2\frac{\xi_{cut}^2 S}{Q^2}
+\log\frac{\xi_{cut}^2 S}{Q^2}\log\frac{k_j\cdot k_i}{2E_j E_i}
-{\rm Li}_2\left(\frac{k_j\cdot k_i}{2E_j E_i}\right)
\nonumber \\*&&\phantom{\frac{1}{8\pi^2}}
+\frac{1}{2}\log^2\frac{2k_j\cdot k_i}{E_j E_i}
-\log\left(4-\frac{2 k_j\cdot k_i}{E_j E_i}\right)
\log\frac{k_j\cdot k_i}{2E_j E_i}-2\log^2 2\Bigg]\phantom{aaaaa}
\label{Iijreg}
\eeqn
and
\beqn
{\cal Q}(\FLNmofullj)&=&\sum_{j=3}^{N+1}\Bigg[\gamma^\prime(a_j)
-\log\frac{S\delta_o}{2Q^2}\left(\gamma(a_j)
-2C(a_j)\log\frac{2E_j}{\xi_{cut}\sqrt{S}}\right)
\nonumber \\*&&\phantom{\sum_{j=3}^{N+1}}
+2C(a_j)\left(\log^2\frac{2E_j}{\sqrt{S}}-\log^2\xi_{cut}\right)
-2\gamma(a_j)\log\frac{2E_j}{\sqrt{S}}\Bigg]
\nonumber \\*&-&
\log\frac{\mu^2}{Q^2}
\Bigg(\gamma(a_1)+2C(a_1)\log\xi_{cut}
+\gamma(a_2)+2C(a_2)\log\xi_{cut}\Bigg).\phantom{aaaa}
\label{Qdef}
\eeqn
Here $E_i$ is the energy of parton $i$ in the partonic center-of-mass
frame, $\mu$ is the factorization scale, and $Q$ is the Ellis-Sexton scale
(see below). The expressions for the colour factors $C(a)$, $\gamma(a)$ and
$\gamma^\prime(a)$ are
\beqn
&&C(g)=\CA\,,\;\;\;\;\;\; C(q)=\CF\,,
\\
&&\gamma(g)=\frac{11\CA-4\TF N_f}{6}\,,\;\;\;\;\;\;
\gamma(q)=\frac{3}{2}\CF\,,
\\
&&\gamma^\prime(g)=\frac{67}{9}\CA-\frac{2\pi^2}{3}\CA
-\frac{23}{9}\TF N_{f}\,,\;\;\;\;\;\;
\gamma^\prime(q)=\frac{13}{2}\CF-\frac{2\pi^2}{3}\CF\,.
\phantom{aaaa}
\eeqn
Notice that eq.~(\ref{Qdef}) depends upon the flavours of the 
initial state partons. In the case of photon-hadron collisions (DIS), we
have $a_1=\gamma$ ($a_1=e$), and in the case of $e^+e^-$ collisions
we have $a_1=e^+$, $a_2=e^-$. Eq.~(\ref{Qdef}) still holds, and we 
just define
\beq
C(a)=0\,,\;\;\;\;\gamma(a)=0\;\;\;\;\;\;{\rm if}\;\;\;\;\;\;
a=\gamma,e\,.
\eeq
We also defined
\beq
d\sigma^{(0)}(\FLNmofullj)=
\uoNmofct\MNmo(\FLNmofullj){\cal S}_{N-1}d\phi_{N-1},
\label{bornfldef}
\eeq
which is identical to eq.~(\ref{borndef}) except for the fact that the 
sum over the flavour of final state partons is not performed, and
\beq
d\sigma_{ij}^{(0)}(\FLNmofullj)=
\uoNmofct\MNmoij(\FLNmofullj){\cal S}_{N-1}\,d\phi_{N-1}.
\label{bornmndef}
\eeq
The functions $\MNmoij$ are usually denoted as colour-linked Born squared
amplitudes. They enter the expression of the virtual corrections to the
$(N-1)$-parton processes, eq.~(\ref{virtdef}). We adopt the following
form~[\ref{ES},\ref{KS}]
\beqn
\MNmoV&=&\frac{\as}{2\pi}\SVfact\Bigg[
-\Bigg(\frac{1}{\ep^2}\sum_{i=1}^{N+1}C(a_i)
+\frac{1}{\ep}\sum_{i=1}^{N+1}\gamma(a_i)\Bigg)\MNmo
\nonumber \\*&&\phantom{\frac{\as}{2\pi}aaaa}
+\frac{1}{2\ep}\sum_{\stackrel{i,j=1}{i\neq j}}^{N+1}
\log\frac{2k_i\cdot k_j}{Q^2}\frac{1}{8\pi^2}\MNmoij
+\MNmoV_{\sss NS}\Bigg].
\label{virtstruct}
\eeqn
This equation also defines the finite part of the virtual correction,
$\MNmoV_{\sss NS}$, which enters eq.~(\ref{sigNmoV}).

The second term in the RHS of eq.~(\ref{Nmobodyxsec}) is
\beqn
d\hat{\sigma}_{\aoat}^{(1,N-1r)}&=&
\frac{\as}{2\pi}\sum_{d}\Bigg\{\xi P_{da_1}^{<}(1-\xi,0)
\Bigg[\uoxic\log\frac{S\delta_{\sss I}}{2\mu^2}
+2\uoxilc\Bigg]
\nonumber \\*&&\phantom{\frac{\as}{2\pi}\sum_{d}}
-\xi P_{da_1}^{\prime <}(1-\xi,0)\uoxic 
-K_{da_1}(1-\xi)\Bigg\}\,{\cal C}_{da_1}
\nonumber \\*&&\times
\FLNmosum d\sigma^{(0)}(d,a_2,\FLNmoj;(1-\xi)k_1,k_2)\,d\xi
\nonumber \\*&+&
\frac{\as}{2\pi}\sum_{d}\Bigg\{\xi P_{da_2}^{<}(1-\xi,0)
\Bigg[\uoxic\log\frac{S\delta_{\sss I}}{2\mu^2}
+2\uoxilc\Bigg]
\nonumber \\*&&\phantom{\frac{\as}{2\pi}\sum_{d}}
-\xi P_{da_2}^{\prime <}(1-\xi,0)\uoxic 
-K_{da_2}(1-\xi)\Bigg\}\,{\cal C}_{da_2}
\nonumber \\*&&\times
\FLNmosum d\sigma^{(0)}(a_1,d,\FLNmoj;k_1,(1-\xi)k_2)\,d\xi\,,
\label{sigNmoR}
\eeqn
where $P_{ab}^{<}(z,0)+\ep P_{ab}^{\prime <}(z,0)+{\cal O}(\ep^2)$
is the Altarelli-Parisi kernel for $z<1$ in $4-2\ep$ dimensions,
the sum $\sum_d$ runs over $g,u,\bar{u},...$, $K_{ab}$ define the finite 
part of the initial state collinear subtraction (in the $\MSB$ scheme 
they are equal to zero), ${\cal C}_{ab}=1$ when $b$ is a quark or a gluon
and, analogously to eq.~(\ref{uoxidef}),
\beq
\int_0^1 d\xi f(\xi)\uoxilc = \int_0^1 d\xi
\Bigg[f(\xi)-f(0)\Th(\xi_{cut}-\xi)\Bigg]\frac{\log\xi}{\xi}\,.
\eeq
In the case of photon-hadron collisions, we define
\beq
P_{d\gamma}(z)=\delta_{dq}\frac{N_{\sss C}}{\TF}P_{qg}(z),\;\;\;\;\;\;
{\cal C}_{d\gamma}=\delta_{dq}e_q^2\frac{\alpha_{em}}{\as}\,.
\eeq
The functions $K_{d\gamma}$ depend upon the scheme adopted for the
partonic densities in the photon, entering the hadronic photon-hadron
cross section (for details on this issue, see for example 
ref.~[\ref{phscheme}]). Finally, $P_{de}=K_{de}=0$, ${\cal C}_{de}=1$.
This formal statement corresponds to the fact that no collinear singularity
is associated with an incoming lepton.

\end{document}